\documentclass[twocolumn,pra,showpacs,superscriptaddress,amssymb,amsmath,amsmath,floatfix]{revtex4-2}
\usepackage[T1]{fontenc}
\usepackage{color,graphicx}
\usepackage{epstopdf}
\usepackage{bm}
\usepackage[colorlinks=true,linkcolor=blue,urlcolor=blue,citecolor=blue]{hyperref}
\usepackage{float}
\usepackage{times}
\usepackage{amsmath, amsthm, amssymb, amsfonts}
\usepackage{xcolor}
\usepackage{orcidlink}

\begin{document}
\title{Dynamics of a Mobile Ion in a Bose-Einstein Condensate}

\author{Piotr Wysocki\orcidlink{0009-0009-1589-1524}}
\affiliation{Faculty of Physics, University of Warsaw, Pasteura 5, 02-093 Warsaw, Poland}
\author{Marek Tylutki\orcidlink{0000-0002-4243-3803}}
\email{marek.tylutki@pwr.edu.pl}
\affiliation{Institute of Theoretical Physics, Wroc{\l}aw University of Science and Technology, 50-370 Wroc{\l}aw, Poland}
\affiliation{Faculty of Physics, Warsaw University of Technology, Koszykowa 75, 00-662 Warsaw, Poland}
\author{Krzysztof Jachymski\orcidlink{0000-0002-9080-0989}}
\email{krzysztof.jachymski@fuw.edu.pl}
\affiliation{Faculty of Physics, University of Warsaw, Pasteura 5, 02-093 Warsaw, Poland}

\date{\today}

\begin{abstract}

Characterization of the dynamics of an impurity immersed in a quantum medium is vital for fundamental understanding of matter as well as applications in modern day quantum technologies. The case of long-ranged interactions is of particular importance here, as it opens the possibility to leverage quantum correlations in controlling the system properties. Here, we consider a charged impurity moving in a bosonic gas and study its properties out of equilibrium. We extract the stationary momentum of the ion at long times, which is nonzero due to the superfluid nature of the medium, and the effective mass which stems from dressing the impurity with the host atoms. The nonlinear evolution leads not only to emission of density waves, but also momentum transfer back to the ion, resulting in the possibility of oscillatory dynamics. 
\end{abstract}

\maketitle
\section{Introduction}
\begin{figure}
	\includegraphics[width = .98\columnwidth]{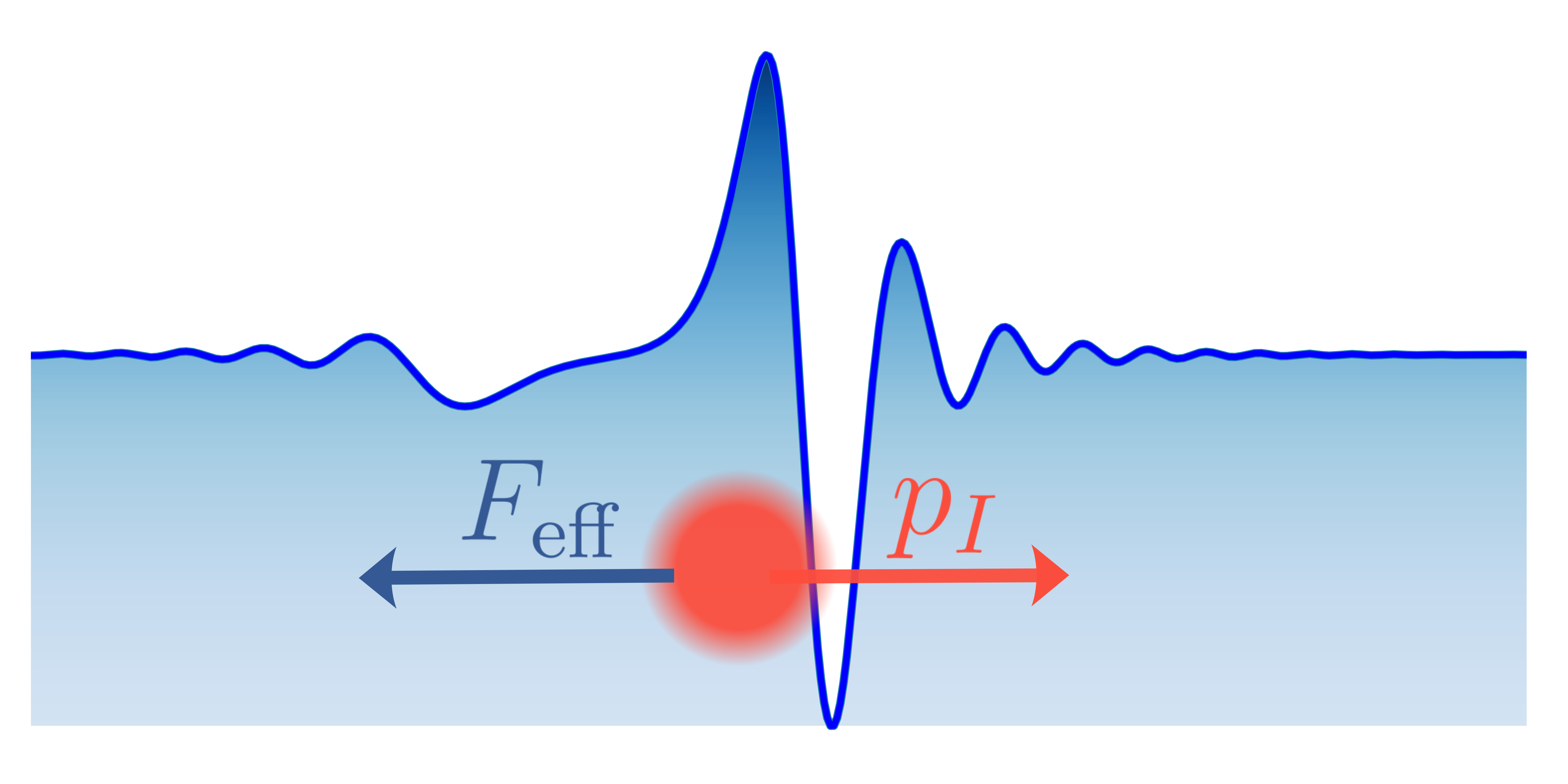}
	\caption{\label{fig:artist} Schematic representation of the studied system of the ion (red circle) interacting with the background BEC (blue background density). The ionic impurity interacts with the condensate and gets dressed by this interaction forming a polaron. Due to the interactions its motion is damped by an effective force (blue arrow), transferring the initial ion's momentum (red arrow) to the condensate. In turn, the extra enery of the condensate is emitted in the form of density perturbations.}
\end{figure}
The conceptually simple problem of injecting an impurity into a medium is fundamental for microscopic understanding of a multitude of phenomena, from condensed matter physics~\cite{Alexandrov2008} to biochemistry~\cite{Stein1986}. It remains relevant on the classical level, where hydrodynamic description may be appropriate, as well as in the quantum regime. In the latter case, a drastic simplification of the problem can be achieved by introducing the concept of a polaron quasiparticle~\cite{Landau1948,Frohlich1954} which is dressed by the medium excitations. While originally introduced in the context of electrons in a crystal lattice, polarons naturally emerge in generic impurity problems. They may serve as a building block for characterizing quantum many-body systems composed of two or more species, but also as a tool introduced for studying a strongly correlated medium or even a sensor of its properties such as the temperature~\cite{Mehboudi2019,Wasak2025}. Quantum impurities attract a lot of attention in the field of ultracold gases, where the system properties can be precisely controlled and one has access to a wide range of measurement techniques~\cite{Grusdt2024,Massignan2025}. Here, one can use a mixture of two atomic species or different internal states of one element with very large imbalance to realize various scenarios. In order to increase the strength and range of the impurity-medium interactions, it is possible to work with a hybrid system of a single ion in an atomic gas~\cite{Tomza2019}, which naturally leads to strong coupling as well as the emergence of mesoscopic many-body bound states~\cite{Astrakharchik2021,Christensen2021}, or Rydberg excitations~\cite{Camargo2018,Engel2024}.

A particularly interesting topic is the nonequilibrium dynamics of polarons, for instance their formation. In a typical experimental situation, the impurity is created from the host gas e.g. by changing the hyperfine state of a handful of atoms~\cite{Skou2021,Yan2024}, or by injecting it into the medium~\cite{Kleinbach2018,Dieterle2021,Albrechtsen2023,Albrechtsen2025}. Characterization of such processes theoretically is much harder than accessing the ground state properties of the system, but vital for understanding quantum transport and solvation dynamics. A similarly challenging topic is the sympathetic cooling of an impurity which is initially much hotter than the medium~\cite{Krych2015,Oghittu2024}. In the presence of strong interactions, the problem is inherently complex and computational approaches are limited to narrow parameter regimes. For instance, in one dimension and for small particle numbers one can make use of tensor network or other numerically exact methods~\cite{Mathy2012,Knap2014,Burovski2014,Grusdt2017,Schurer2015,Schurer2017,Mistakidis2019}. However, even these are computationally very expensive and in general some simplifying assumptions are necessary. 

A well established theoretical approach towards polaron problems starts with a transformation to the reference frame co-moving with the impurity~\cite{LLP,Gross1962}. This is useful especially if the total momentum of the system is vanishing and allows for elimination of the impurity degrees of freedom. For the case of the Bose polaron, i.e. when the medium is bosonic, one can then assume a gaussian state of Bogoliubov excitations~\cite{Shchadilova2016,Drescher2019,Seetharam2021,Seetharam2024,Yougurt2025} or proceed with a mean field ansatz which leads to a modified Gross-Pitaevskii equation (GPE)~\cite{Mistakidis2019,Jager2020,Hryhorchak2020,Guenther2021,Cavazos2024} expected to work well for small gas parameters~\cite{Yegovtsev2024}. 

In this work, we develop a convenient treatment of a mobile impurity moving in a quantum degenerate bosonic medium in the presence of  long-ranged interactions, particularly relevant for the case of a charged impurity. Working in the co-moving frame, we apply the mean field ansatz to the Bose gas, which is interpreted as equilibration of the condensate in the vicinity of the ion due to the strong coupling. This approach allows for straightforward calculation of the dynamics, including different types of initial conditions. The effective mass of the polaron can be computed by fitting the energy dispersion relation to a free quasiparticle at low momenta. We characterize the polaron formation and cooling timescales and find that they strongly depend on the initial state of the system. Interestingly, for very strong interactions the nonlinear dynamics results in oscillations of the impurity momentum, regardless of the system's dimension. 

The paper is organized as follows. In section~\ref{sec:system} we formulate the problem and describe the method. Then in section~\ref{sec:results} we present the calculations for one- and three-dimensional systems. We discuss the  results and their possible implications in~\ref{sec:disc} and conclude the work in section~\ref{sec:concl}.

\section{The system}
\label{sec:system}
We consider a single ion as an impurity immersed in a gas of bosonic atoms, described by the many-body Hamiltonian
\begin{equation}
\label{eq.mbham}
\mathcal{H} = \frac{{\bm p}_I^2}{2m_I}+\sum_i \frac{{\bm p}_i^2}{2m_B}+\sum_i V({\bm x}_i - {\bm x}_I) + \sum_{i<j} U({\bm x}_i - {\bm x}_j) ~,
\end{equation}
where ${\bm p}_i$ are the momenta of the atoms forming the background gas, ${\bm p}_I$ is the momentum of the impurity, and $m_B$, $m_I$ are the atomic masses of the respective species. The two-body interaction between the bosonic atoms is denoted as $U({\bm x}_i - {\bm x}_j)$ and between the ion and a background boson as $V({\bm x}_i - {\bm x}_I)$, for which we employ a regularized potential~\cite{Krych2015}
\begin{equation}
\label{eq.ionpot}
V({\bm x}) = -\frac{C_4}{({\bm x}^2 + b^2)^2}
\end{equation}
with a tunable parameter $b$ and strength $C_4$ proportional to atomic polarizability, while the bosons are assumed to interact with a short-range pseudopotential described by $U({\bm x}_i - {\bm x}_j) = g\delta(\mathbf{r}_i-\mathbf{r}_j)$.
We now perform a unitary Lee-Low-Pines (LLP) transformation to the frame co-moving with the impurity~\cite{LLP,Girardeau1961,Drescher2020} generated by $S = {\bm x}_I \cdot \sum_i {\bm p}_i$. Then in this frame we make the mean field assumption for bosons, assuming that due to the strong ion-atom interaction the gas rapidly equilibrates around the moving ion. It is important to note, that the mean-field ansatz assumes that number of particles in the condensate is large enough to be well approximated with the thermodynamic limit. This in turn ensures that the finite size and many-body corrections are negligible, and the LLP transformation for uniform systems is applicable. This leads to an energy functional of a simple form
\begin{equation} \label{eq.edf}
   \mathcal{E}[\psi] = \mathcal{E}_0[\psi] + \frac{({\bm p}_0 - \langle {\bm P} \rangle)^2}{2m_I}\\
\end{equation}
where $\mathcal{E}_0$ is the energy of the standard Gross-Pitaevskii functional 
\begin{equation*}
   \mathcal{E}_0[\psi] = \int d{\bm x} \left( \frac{\hbar^2}{2m_r}| \nabla \psi|^2 + V({\bm x}) |\psi|^2 + \frac{g}{2} |\psi|^4 \right)
\end{equation*}
but with the reduced mass, $m_r = m_I m_b / (m_I + m_b)$. The initial total momentum of the system, which is a conserved quantity due to the translational symmetry, is denoted by ${\bm p}_0$. The expectation value of the condensate momentum yields
\begin{equation*}
\langle {\bm P} \rangle \equiv {\bm p}_0 - {\bm p}_I = -i \hbar \int d{\bm x} \, \psi({\bm x}, t)^* \nabla \psi({\bm x}, t) ~.
\end{equation*}
The variation of Eq.~(\ref{eq.edf}) with respect to the condensate's wavefunction leads to a modified GPE that reads
\begin{equation}\label{eq:gpe}
   i \hbar \frac{\partial \psi}{\partial t} = -\frac{\hbar^2}{2m_r} \nabla^2 \psi + V({\bm x}) \psi + g |\psi|^2 \psi + \frac{i\hbar}{m_I} {\bm p}_I \cdot \nabla \psi ~,
\end{equation}
with the additional last term that stems from the LLP transformation. This term describes the total momentum conservation and disappears (i) for an infinitely heavy ion, i.e. $m_I \to \infty$, or (ii) when both the ion and atoms have zero momentum. We solve Eq.~\eqref{eq:gpe} numerically (more details are provided in the Appendix). 

The problem features multiple length and energy scales, which makes the dynamics complex. Here, we employ the condensate healing length $\xi=\hbar/\sqrt{m_b n_0 g}$ with $n_0$ being the asymptotic gas density away from the ion as the unit of length, and measure energies in units of $\varepsilon=n_0g$, which is the corresponding chemical potential. Normalization of the wave function sets the total number of bosons via $\int d {\bm x} \, |\psi({\bm x})|^2 = N_b$. The ion-atom interaction introduces an additional length scale $R^\star=(2m_r C_4/\hbar^2)^{1/2}$ and energy scale $E^\star=\hbar^2/(2m_r R^\star)^2$. Further parameters include the ion's and atoms' masses and the strength of interatomic interaction $g$, which can be related to the scattering length. Note that in this set of units time is measured in $\tau=\hbar/n_0g$, $g = (\xi^d n_0)^{-1}$ with $d$ being the dimensionality of the system, and the speed of sound $c=1$. Finally, we set $b = 1.5$, which provides a slowly varying potential that does not increase the gas parameter in the center beyond the validity of the mean field approximation. The interaction is attractive and rather weak, such that for $C_4 = 1$ in the three-dimensional case the two-body system does not support a bound state. Nevertheless, as we will demonstrate the long-range nature of the potential leads to rich impurity dynamics.

\section{Results}
\label{sec:results}
We study Eq.~\eqref{eq:gpe} aiming both for its stationary solutions and time dynamics. From the obtained wave function, we calculate observables such as the density distribution and the total energy as a function of the initial momentum. If the system forms a massive quasiparticle, the dispersion relation is expected to be quadratic and we can extract the effective mass of the dressed impurity (polaron) as $E({\bm p}_0)=E_0 + {\bm p}_0^2 / (2m^\star)$. 

In order to get a better understanding of the dynamics of the impurity interacting with the condensate, we derived an analytical equation describing the motion of the impurity, introducing an effective damping force stemming from the interaction with the background condensate, 
\begin{equation*}
	{\bm F}_{\rm eff} = \frac{d{\bm p}_I}{dt} = i \int d{\bm x} (\dot{\psi}^* \nabla \psi + \psi^* \nabla \dot{\psi}) ~,
\end{equation*}
which leads to (see Methods)
\begin{equation} \label{eq.dpdt}
	{\bm F}_{\rm eff} = \int d{\bm x} \, \nabla V({\bm x}) \delta \rho({\bm x}) ~,
\end{equation}
where $\delta \rho$ is the condensate's density fluctuation with respect to the initial, symmetric density, $n({\bm x}) = n_0({\bm x}) + \delta \rho({\bm x})$. We checked numerically that this relation is indeed exact, see e.g. Fig.~\ref{fig:1dflutter}(b), which also in turn shows that our numerical solutions are free of finite size effects. Similar drag force term has also been introduced in the context of superfluid flow past an obstacle~\cite{Pavloff2002}.

It has been suggested that the asymptotic impurity momentum in one dimension depends on the details of the protocol~\cite{Gamayun2018}. We therefore consider two scenarios for the initial state of the dynamics: (i) a rapid quench, which consists of a sudden introduction of a moving ion to a previously homogeneous BEC, and (ii) initializing the dynamics from the staionary solution of the system, where the ion is at rest, and subsequently giving it a momentum ${\bm p}_0$. The simplicity of our approach allows us to study the nonequilibrium ion dynamics in any dimension. Below, we will discuss both one-dimensional (1D) and three-dimensional (3D) scenarios. 

\subsection{One dimension}
\begin{figure}
\includegraphics[width = .98\columnwidth]{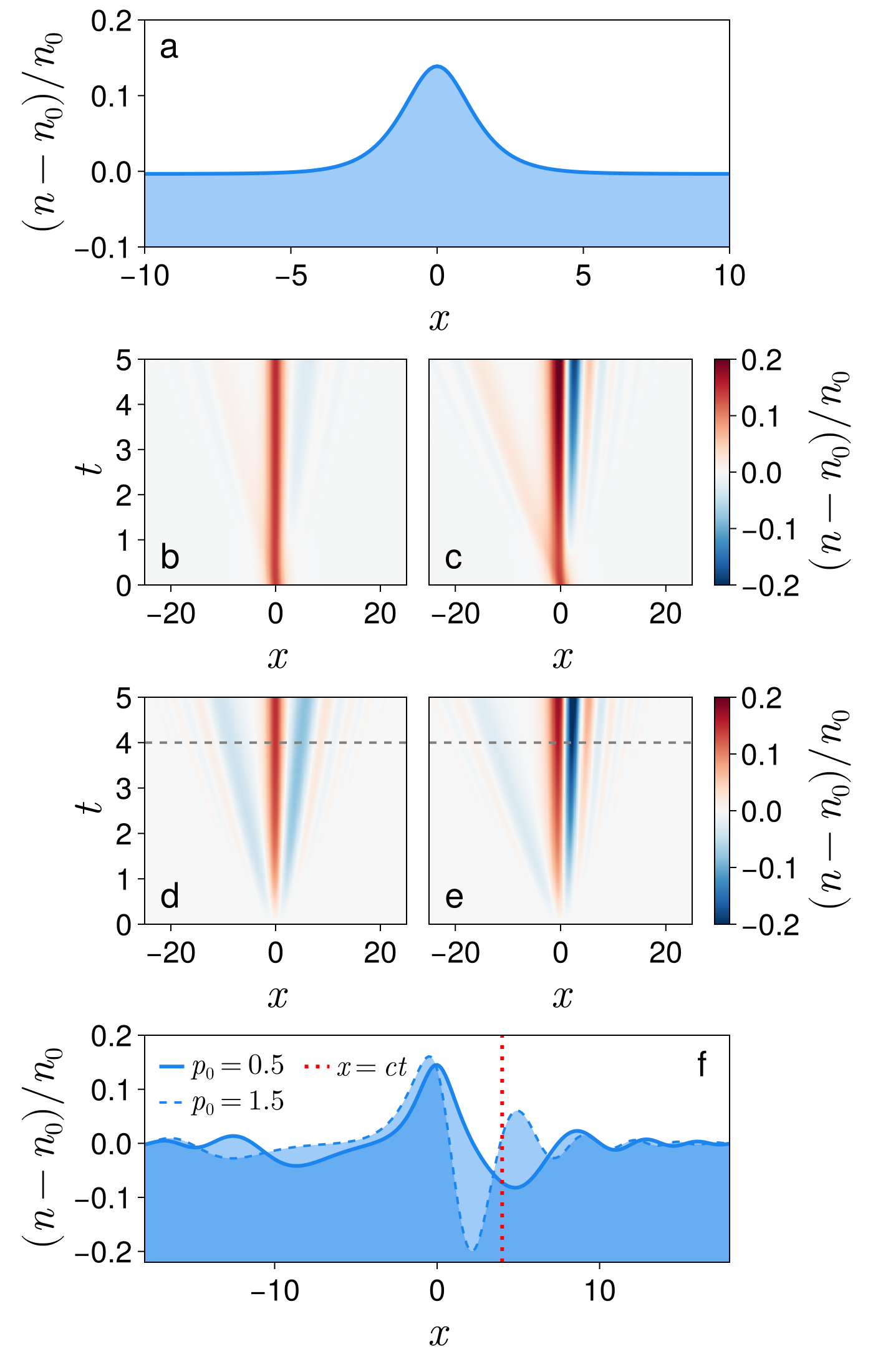}
	\caption{\label{fig:1ddensity} Summary of the static and dynamical solutions for the 1D system. (a) Ground state density profile of the Bose gas interacting with a stationary charged impurity (the interaction strength is set to $C_4 = 1.0$); (b,c) space-time diagram showing the time evolution of the condensate density, where the initial state corresponds to that from panel (a) and for $|{\bm p}_0| = 0.5$, $|{\bm p}_0| = 1.5$ respectively; (d,e) space-time diagram showing the time evolution of the condensate density after a quench from the noninteracting system, for the same momenta as in panels (b,c); (f) a cut through the BEC density at time $t = 4$ for the case of the noninteracting protocol shown in panels (d,e), the red dashed line shows the propagation of sound with the speed of $c = 1$. }
\end{figure}
\begin{figure}
\includegraphics[width = .98\columnwidth]{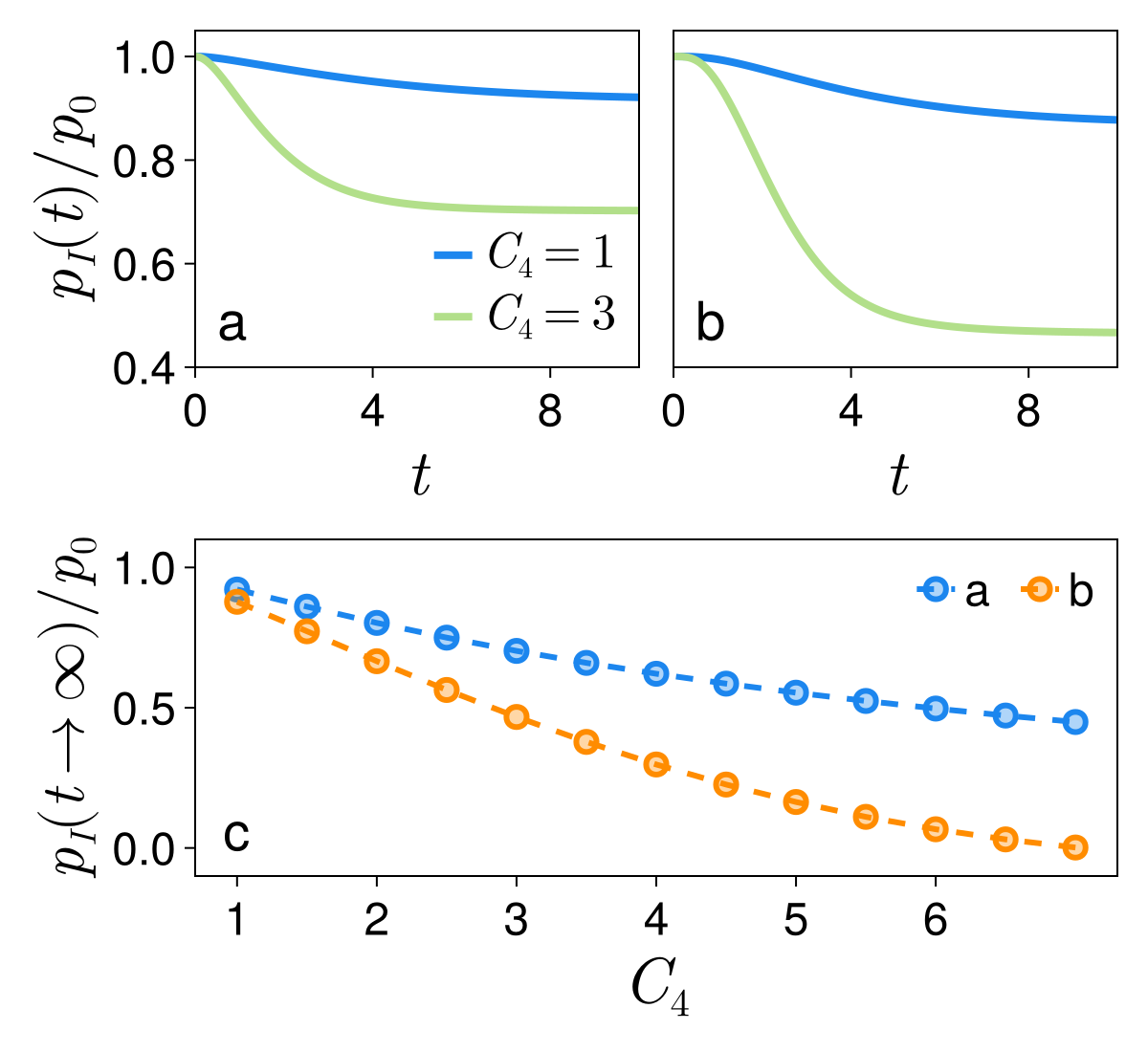}
	\caption{\label{fig:1dmomentum}The ion momentum as a function of time for (a) a quench from static ion and (b) a quench from non-interacting system. (c) Asymptotic ion momentum ${\bm p}_I(t \to \infty)$ as a function of the ion-BEC interaction strength for the same two quench protocols. Here, we present results for $|{\bm p}_0| = 1$ in our units. }
\end{figure}
\begin{figure}
\includegraphics[width = .98\columnwidth]{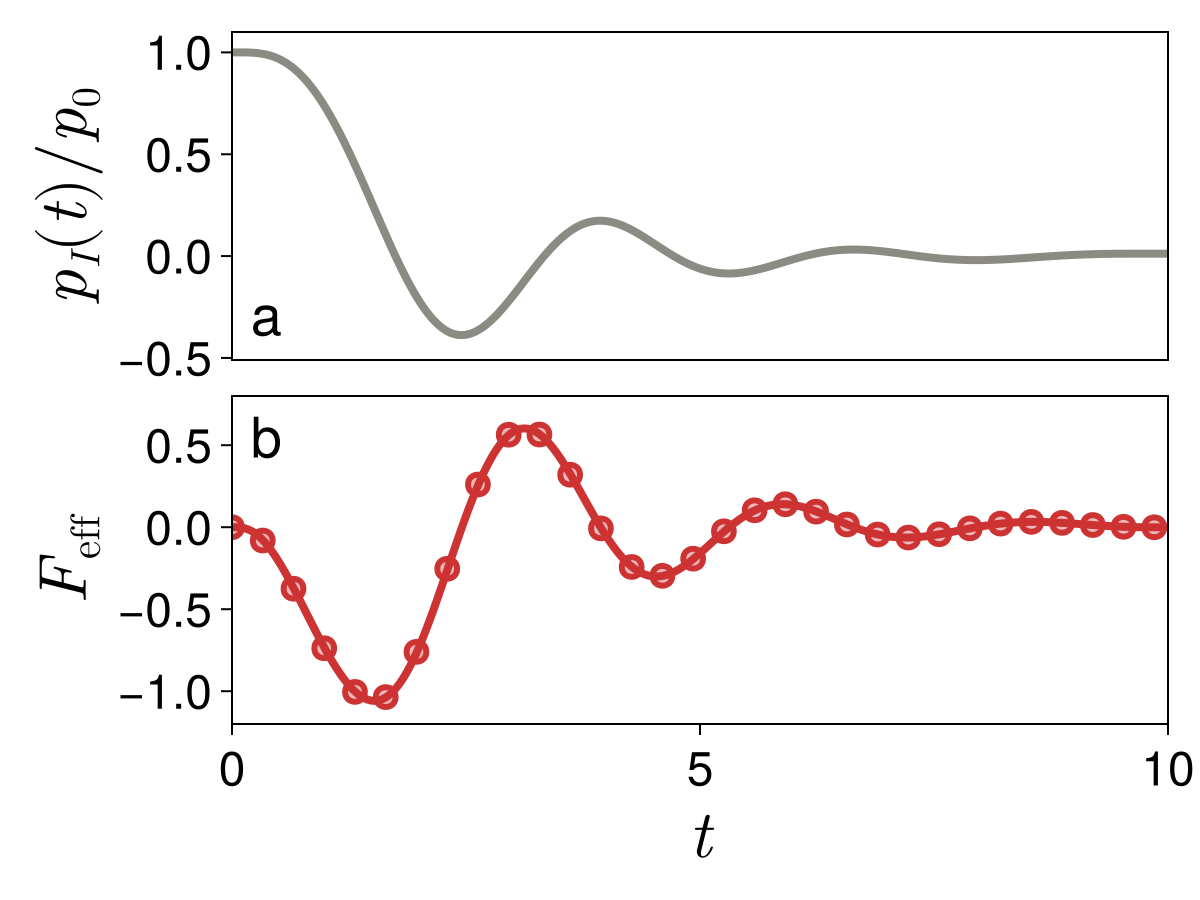}
	\caption{\label{fig:1dflutter} (a) An example of a ``quantum flutter'', when strong interactions lead to oscillations of the ion's momentum ($C_4 = 6$ and $g = 0.1$). (b) The effective damping force acting on the impurity calculated from the numerical results (solid line) and from Eq.~(\ref{eq.dpdt}) (circles). Here, $|{\bm p}_0| = 1$; the effect is largely independent of the initial momentum.}
\end{figure}
\begin{figure}
\includegraphics[width = .98\columnwidth]{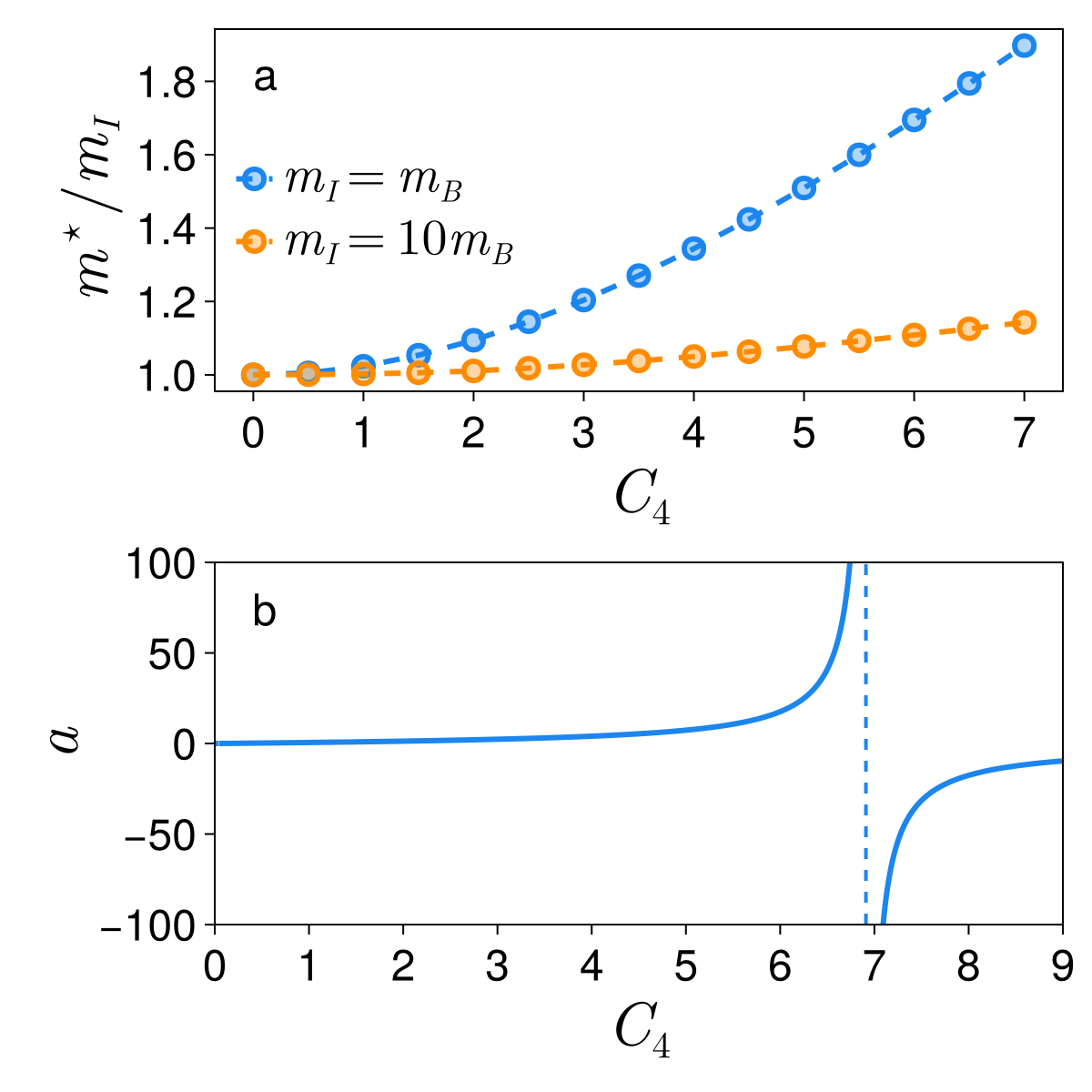}
	\caption{\label{fig:1dmass} (a) Effective mass for a light ion ($m_I = m_B$, blue circles) and a heavy ion ($m_I = 10\, m_B$, orange circles) as a function of the interaction strength $C_4$ in 1D. The dashed lines are added as a guide for the eye. (b) Scattering length corresponding to the one-dimensional version of the potential \eqref{eq.ionpot} as a function of the interaction strength $C_4$ in a wider range, showing the position of the scattering resonance.}
\end{figure}
We first present our results for the dynamics of a 1D system for the two scenarios discussed above (interaction quench and initially static configuration). We also consider the different ion mass ($m_I$, light and heavy ions) and different densities of the BEC (dilute and more dense Bose gas). We first choose $C_4 = 1.0$ in Eq.~(\ref{eq.ionpot}) as the interaction strength, which means that at the distance of one healing length away from the impurity the interaction is still equal to $n_0 g$.
Following Ref.~\cite{Catani2012}, we choose the particle number $N_B = 200$ and the density $n_0 \approx 1.5 \xi^{-1}$, and $g \approx 0.66$. In all simulations we set $m_I = m_B$, except for effective mass calculations, where we compare this case with a heavy ion $m_I = 10\, m_B$, relevant for ongoing experiments, in which large mass ratio is beneficial for sympathetic cooling and allows for reaching the regime where Feshbach resonances can be used to tune the interactions~\cite{Tomza2019,Weckesser2021}.

In Fig.~\ref{fig:1ddensity}(a) we show the density profile of the condensate interacting with the stationary ion (${\bm p}_0 = 0$). In spite of a weak interaction, the density around the ion increases by about 20\%. The dynamics of the condensate after the quench procedure is shown in space-time plots in panels Fig.~\ref{fig:1ddensity}(b,c) for the momentum-kick procedure that starts with a stationary solution of an interacting system and in panels Fig.~\ref{fig:1ddensity}(d,e) for the interaction quench that starts from the non-interacting system. The interaction of the condensate with the moving ion clearly introduces a gradual asymmetry in the condensate's density, causing the momentum transfer between the impurity and the gas, as explained by the Eq.~(\ref{eq.dpdt}). This can be interpreted as emission of phonons, which carry away some of the momentum. The speed of propagation of these density fluctuations is roughly consistent with the speed of sound, as shown in Fig.~\ref{fig:1ddensity}(f) by the red dashed veritcal line. However, note that due to the increased gas density the local speed of sound around the impurity is larger than one, which makes the initial dynamics faster.

The impurity exchanges momentum with the gas, leading to deceleration and ultimately a stationary state is expected to be formed. As the medium is superfluid, there is no damping force at zero temperature and the asymptotic ion momentum does not need to vanish. Figure~\ref{fig:1dmomentum}(a,b) shows the time evolution of the impurity momentum, which indeed stabilizes, and its asymptotic value is summarized in Fig.~\ref{fig:1dmomentum}(c) for different interaction strengths. Note that the stationary state depends on the initial condition, and the asymptotic momentum is generally larger when starting from the polaronic ground state than from the homogeneous gas. This is intuitively clear, as in the latter case the evolution is more complex due to the competition of dragging the bosons to the vicinity of the ion and moving of the ion through the gas. 

As the ion-BEC coupling is increased further relatively to the strength of the contact interaction between the bosons, another nontrivial effect starts manifesting. Namely, we observe damped oscillations of the ion momentum, which get stronger as the gas becomes more compressible. The periodic transfer of momentum between the ion and the condensate, shown in Fig.~\ref{fig:1dflutter}(a), resembles breathing dynamics or quantum flutter observed in one dimension for strongly interacting supersonic impurities~\cite{Mathy2012,Knap2014}. In our case, we even observe that the momentum of the impurity can assume negative values at times, in contrast to the intuition from classical hydrodynamics. In panel Fig.~\ref{fig:1dflutter}(b) we calculate the damping force and compare it with Eq.~(\ref{eq.dpdt}). Perfect agreement indicates that this effect is an inherent feature of the nonlinear extended GPE equation.  

Finally, we extract the mass of the polaron $m^\star$ (see Methods), as the bare mass of the ion gets renormalized by the interaction with the condensate. Figure~\ref{fig:1dmass} shows the renormalized mass in units of the bare mass as a function of the interaction strength $C_4$. We show our results for both the light ion ($m_I = m_B$) and for the heavy ion ($m_I = 10 m_B$). In agreement with intuition, the inertia of the heavy ion results in a smaller effect of the interaction on the effective mass. We note that the obtained mass increase is of the order of one, which indicates that we are indeed dealing with a polaronic quasiparticle rather than a many-body molecular bound state, which would feature much larger effective mass~\cite{Astrakharchik2021}. In addition, we show in panel b the dependence of one-dimensional scattering length on $C_4$, which indicates that the resonance occurs for $C_4$ larger than 7. For $C_4$ equal to one, the scattering length $a_{1D}\approx 0.5$ is comparable to the range of the potential as well as to the healing length.

\subsection{Three dimensions}
\begin{figure}
\includegraphics[width = .98\columnwidth]{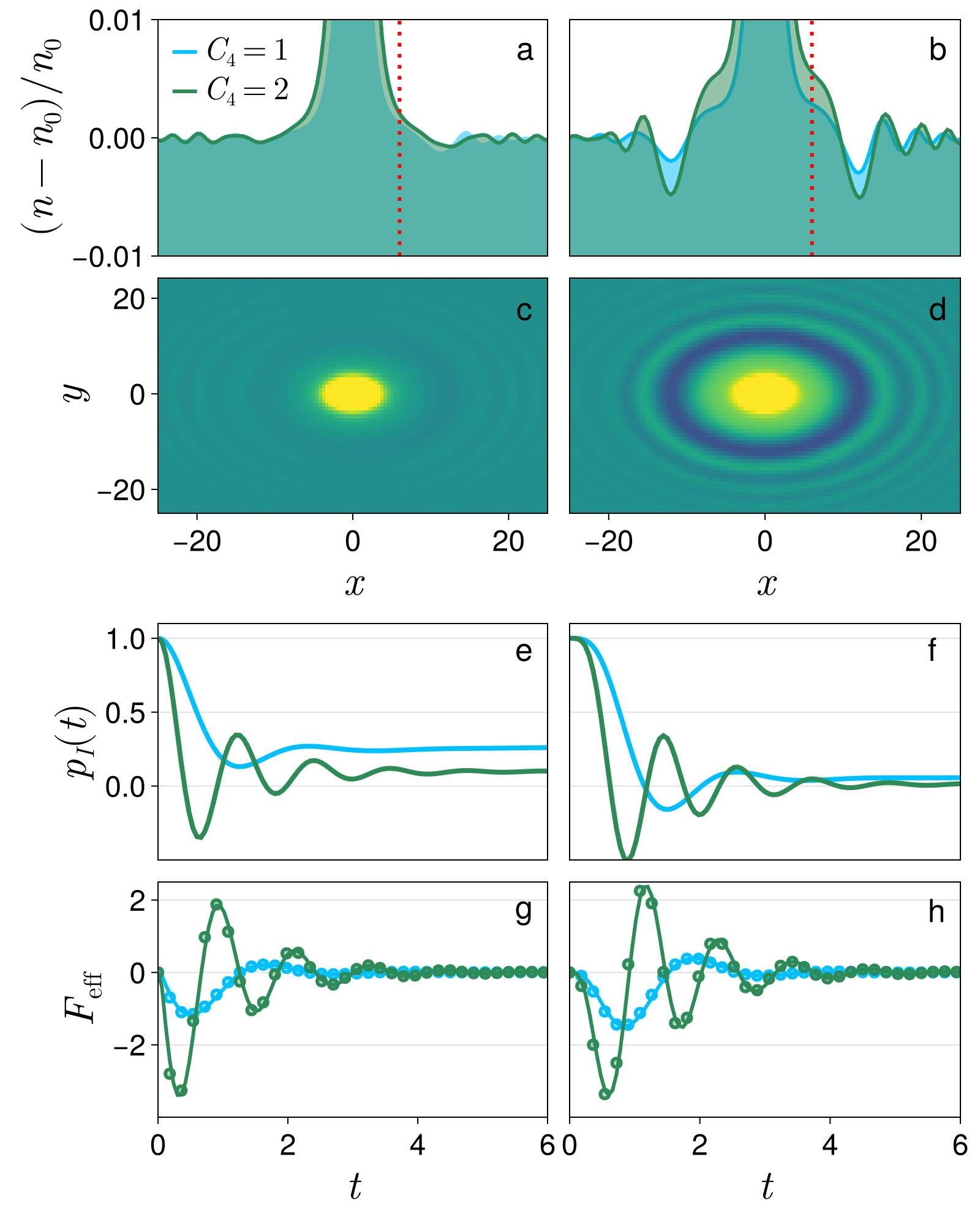}
	\caption{\label{fig:2ddensity}
	One-dimensional cuts of the density profiles of the 3D condensate at $t = 10$ for two different interaction strength values $C_4$ after (a) the quench from the static solution (${\bm p}_0 = 0$) of the interacting system and (b) the quench from the non-interacting system. The red dashed line corresponds to $x = ct$ showing the speed of sound for comparison. The corresponding two-dimensional cuts are shown in (c) and (d) for the $C_4 = 2$ case (color scale range same as $y$-axis range above). The momentum dynamics is shown in (e) and the damping force in (g) for the quench from the static solution and in (f,h) for the quench from the non-interacting system, respectively. In (g,h), the continuous lines show the results of direct simulations, and the circles show the damping force calculated from Eq.~\eqref{eq.dpdt}
	}
\end{figure}

\begin{figure}
\includegraphics[width = .98\columnwidth]{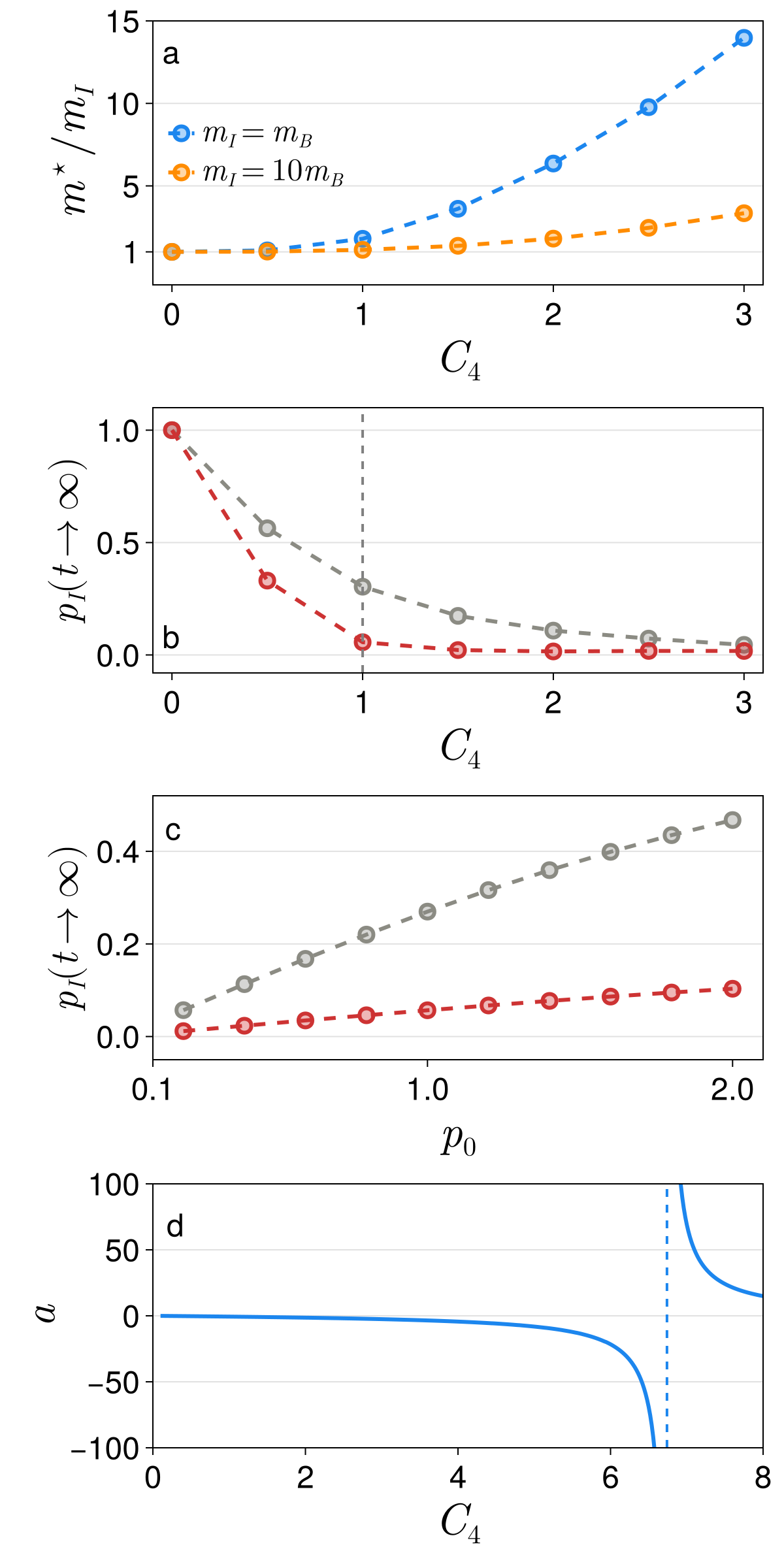}
	\caption{\label{fig:2longtime} Summary of the polaron dynamics in 3D. In (a) we show the effective mass as a function of the interaction strength for the light ($m_I = m_B$, blue circles) and heavy ion ($m_I = 10\, m_B$, orange circles). (b) The final, asymptotic momentum of the ion at $t \to \infty$ after the equilibration process has taken place. The grey data points correspond to the quench from the static ion in the interacting system, while the red data depicts the quench from the noninteracting system. Here we fix the initial momentum at $p_0 = |{\bm p}_0| = 1$. (c) Asymptotic momentum at $t \to \infty$ as a function of the initial momentum $p_0$, where $p_0=1$ marks the speed of sound, for a fixed interaction strength set to $C_4 = 1.0$ (see the dashed, vertical line in (b)). (d) Scattering length corresponding to the potential in Eq.~(\ref{eq.ionpot}) as a function of the interaction strength $C_4$ in a wide range of values. }
\end{figure}
We now turn to simulations in 3D, where the interatomic coupling constant is defined as $g = (\xi^3 n_0)^{-1}$. Following the parameters of the experiment in Ref.~\cite{Jorgensen2016} we set $n_0 \approx 142 \xi^{-3}$, which corresponds to $g \approx 0.007$. In our simulations, the spatial domain is a box of size $100\times50\times50~\xi^3$ (elongated in the direction of the ion's momentum), and we assume $m_I = m_B$. We present the cuts of the density profile of the condensate for quenches from interacting and non-interacting initial states in Fig.~\ref{fig:2ddensity}(a,b). Similarly to 1D, the atractive interaction between the condensate and the impurity creates a region of increased density around the ion, and the asymmetric perturbation propagates as the dynamics unfolds in time. The perturbations propagate with the speed of sound as shown by the vertical red dashed line, which marks the position where a perturbation travelling with a speed of sound would be after the time $t$ corresponding to the data presented in the figure. Then, we calculate the value of the ion momentum $p_I = |{\bm p}_I|$ and observe its dynamics in time. In analogy to the 1D case, the impurity momentum saturates for large times, reaching its equilibrium value. If the interaction strength between the impurity and the condensate is large enough, we once again observe the flutter, where the momentum transfer between the ion and the condensate takes place in both directions, leading to the damped oscillations of the ion momentum, as shown in Fig.~\ref{fig:2ddensity}(e,f). Finally, we calculate the damping force, that the condensate exerts on the impurity and we compare these results with the Eq.~(\ref{eq.dpdt}), as is presented in Fig.~\ref{fig:2ddensity}(g,h), again obtaining perfect agreement. Note that the scattering length is typically negative, indicating the lack of bound states. The first one appears at $C_4>6$, as marked by the resonance.

The large time dynamics and the renormalization of the ion mass is summarized in Fig.~\ref{fig:2longtime}. In analogy to the 1D case, the interactions dress a light ion stronger than a heavy one. For $m_I\to\infty$, the last term in Eq.~\eqref{eq:gpe} becomes irrelevant, and the gas experiences a static potential, while the presence of a mobile (finite mass) impurity induces an additional force. Note that in 3D the effect is much more dramatic than in 1D, as more bosons can reside in the potential well around the ion and take part in a collective motion. The effective mass can be estimated numerically in Monte Carlo calculations by studying the impurity dispersion during imaginary time evolution~\cite{Ardila2015,Ardila2020}, which in the 3D case close to unitarity gives $m^\star\approx 6m_B$~\cite{Astrakharchik2021}, in qualitative agreement with our results. 

In Fig.~\ref{fig:2longtime}(b) we show the asymptotic momentum of the ion, i.e. at $t \to \infty$, after equlibration. As expected, as the interactions become stronger the damping effect caused by phonon emission is more pronounced, and the ion is significantly slowed down. For moderately strong interactions, i.e. $C_4\gtrsim 0.5$ the stationary momentum drops significantly, especially for an initially homogeneous gas where the impurity loses more than 90\% of its initial velocity. The dependence of the final momentum on the initial momentum of the ion is summarized in Fig.~\ref{fig:2longtime}(c). We observe that the dynamics is nontrivial, and there is an optimal initial momentum of the impurity for which it becomes least affected by the damping and for which the motion stabilizes with the largest final velocity.

\section{Discussion}
\label{sec:disc}
The mean velocity of the ion is governed by the force given by Eq.~\eqref{eq.dpdt}. This seemingly simple formula involves an integral of the gradient of the potential with the density fluctuations, meaning that the force has a nonlocal origin. In other words, once an excitation is emitted, it still interacts with the ion due to the finite range of the potential. This is a more complex scenario than for the case of contact interactions, where the superfluid medium is not strongly affected and it is possible to identify shock waves analogous to Cherenkov radiation above the critical velocity~\cite{Seetharam2021}. Here, the interaction decays with the distance as a power law, and so does its gradient, leading to meaningful effects at moderate time scales and making the dynamics nonlinear, as excitations emitted at different times experience ion-mediated interaction. Note also that the effective speed of sound is decreasing with the distance from the ion, further complicating the dynamics as an emitted wave may undergo bending effects due to inhomogeneity. For short-ranged potentials the density modification is more localized and the phonons quickly reach the region where the impurity is irrelevant, so one can expect the dynamics to become stationary much faster. Nevertheless, for very weak interactions Born approximation should hold and the gas density would be homogeneous,making the effective mass universal, i.e. described by the scattering length instead of the full functional form of the interaction potential~\cite{Wysocki2026}. 

The observed oscillations of the ion momentum, along with the possibility of the ion changing the propagation direction, result from the above mentioned nonlinear effects and are not limited to one dimension. In order to emerge, they require sufficiently strong coupling between the impurity and the gas. We note that a qualitatively similar effect has been predicted for the case of a Fermi gas, where the impurity dynamics within the hydrodynamic treatment is also described by a nonlinear differential equation~\cite{Mysliwy2025}. This indicates that the oscillatory behavior is fundamentally different from the quantum flutter~\cite{Mathy2012}, which originates from the specific structure of excitations of 1D bosonic systems~\cite{Zhang2024}. In our case oscillations seem to be generic within the mean field model employed here, and exist for any dimension. Note that oscillations have also been observed numerically by~\cite{Drescher2019} for repulsive interactions and interpreted as the influence of the bound state occupation. However, in our system the interactions are attractive and typically too weak to feature a two-body bound state, pointing to the nonlinear effects as the explanation.

Finally, let us comment on the validity of the mean field approximation for a moving ion. Similarly to other approaches taking advantage of the weak interactions between the bosons, the gas parameter is required to be small everywhere (not only away from the impurity). For this reason we have to resort to regularized interaction potential with a small depth. In this work, the mean field assumption is applied in the reference frame of the ion instead of the laboratory frame, meaning that the condensate is co-moving with the impurity. This intuitively applies better for strong coupling, where the ion-atom potential is stronger than the energy scales associated with the gas. While this approach has been shown to work for the static case~\cite{Yegovtsev2024}, the motion of the ion is expected to induce additional correlations in the host gas, particularly when the impurity is suddenly created and the polaron has yet to be formed. We thus expect our description to work best at low ion momentum, when the gas has more time to adjust, and for heavy impurities. Crucially, the effective mass is extracted by expanding around the static case, making the results rather reliable, and large mass ratio corresponds to the regime of most interest for current experiments~\cite{Weckesser2021,Lous2022}. Note also that in the co-moving frame the speed of sound is modified due to replacing the boson mass with the reduced mass. In one dimension and for contact interactions, keeping the boson mass in the energy functional provides an estimate of the ground state energy at a given momentum which is closer to the analytically derived Yrast line~\cite{Majumdar2025}. Additional benchmarking against other methods, especially in 3D, would be very useful to optimize the mean field approach.

\section{Conclusions}
\label{sec:concl}
We studied the dynamics of a long-range interacting impurity immersed in a Bose gas using mean field approach in the co-moving frame. This method is rather general in scope: it is capable of describing long-ranged interactions as well as the strong coupling regime, includes the density inhomogeneities of the medium, and works in any dimension. We obtained the effective mass of the created polaron and described the impurity motion, including its possible velocity oscillations and the asymptotic momentum, which strongly depends on the initial conditions. 

This method can be extended to more involved scenarios, such as starting with a different kind of an excited state, possibly involving vortices, or making different assumptions about the properties of the medium. Furthermore, it is possible to increase its accuracy by extending the mean field ansatz, e.g. by inlcuding short-range bosonic correlationsvia a Jastrow term~\cite{Drescher2020}.  This could allow for more precise description of the impurity-induced interactions between the bosons.

{\it Note added.} We became aware of a recent work~\cite{Majumdar2025} studying the case of an impurity in a one-dimensional gas with repulsive contact interactions, providing results complementary to ours.

\section*{Acknowledgments}
We are grateful to Michael Knap for inspiring discussions. This work was supported by the National Science Centre, Poland (NCN), Contracts No. 2019/35/D/ST2/00201 (M.T.) and No. 2020/37/B/ST2/00486 (P.W., K.J.)

\section*{Author contributions}
K.J. conceived the project. P.W. performed all numerical simulations and prepared the figures. M.T. derived the equation for the effective damping force. M.T. and K.J. performed approximate examination of the dynamics and wrote the first version of the manuscript. All authors contributed to conceptual work, analysis of the results and writing the final version of the manuscript.

\appendix

\section{Derivation of the  modified GPE and the drag force}

We begin with the many-body Hamiltonian given by the Eq.~(\ref{eq.mbham}). After performin the LLP transformation, we make the mean-field ansatz, and the resulting energy functional assumes the following form~\cite{LLP,Girardeau1961,Drescher2020}
\begin{equation} \label{eq.edf2}
	\mathcal{E} = \mathcal{E}_0 + \frac{{\bm p}_I^2}{2m_I}\\
\end{equation}
and $\mathcal{E}_0$ is the standard mean-field energy of the Gross-Pitaevskii equation in the laboratory frame
\begin{equation*}
	\mathcal{E}_0 = \int d{\bm x} \left( \frac{\hbar^2}{2m_r}| \nabla \psi|^2 + V({\bm x}) |\psi|^2 + \frac{g}{2} |\psi|^4 \right)
\end{equation*}
but with the reduced mass, $m_r = m_I m_b / (m_I + m_b)$. The ion's momentum is ${\bm p}_I \equiv {\bm p}_0 - \langle {\bm P} \rangle$, and the condensate's momentum yields $\langle {\bm P} \rangle = -i \hbar \int d{\bm x} \, \psi({\bm x}, t)^* \nabla \psi({\bm x}, t)$. The variation of Eq.~(\ref{eq.edf2}) leads to the following modified GPE:
\begin{equation}\label{eq:gpe2}
	i \hbar \frac{\partial \psi}{\partial t} = -\frac{\hbar^2}{2m_r} \nabla^2 \psi + V({\bm x}) \psi + g |\psi|^2 \psi + \frac{i\hbar}{m_I} {\bm p}_I \cdot \nabla \psi
\end{equation}
We adopt the following units for length $\xi = \hbar / \sqrt{m_b n_0 g}$, energy $\varepsilon = n_0 g$, time $\tau = \hbar / n_0 g$, and the wave function $1 /\sqrt{\xi}$. The condensate wave function normalized to the number of bosons in the system, $\int d {\bm x} \, |\psi({\bm x})|^2 = N_b$. 

We use the quantity $(\xi^d n_0)^{-1}$ as the mean field interaction coupling, where $d$ is the dimensionality of the system, and the resulting dimensionless GPE reads
\begin{equation*}
	i \frac{\partial \psi}{\partial t} = -\frac{m_b}{2 m_r} \nabla^2 \psi + V({\bm x}) \psi + \frac{1}{\xi^d n_0} |\psi|^2 \psi + i \frac{m_b}{m_I} {\bm p}_I \cdot \nabla \psi
\end{equation*}
The condensate momentum can be conveniently calculated in the Fourier space $\langle {\bm P} \rangle = \int d{\bm k} \, {\bm k} |\varphi({\bm k})|^2 / (2 \pi)^3$ with $\varphi({\bm k}) = \int d{\bm k} \, \psi({\bm x}) e^{-i \bm{k x}}$. Note that the speed of sound in these units is $c = \sqrt{n_0g / m_b} = \xi / \tau = 1$, and the effective mass $m^\star$ is calculated as $1 / m^\star = 2 \lim_{p \to 0} \partial E / \partial (p^2)$, where $E$ is the total energy of the interacting system. We solve Eq.~(\ref{eq:gpe2}) with the Split-Step Fourier method treating the impurity's momentum term in the Fourier space. 

In order to gain insight into the relation between the momentum transfer and condensate dynamics, we derive a differential equation for the impurity's momentum as
\begin{equation}
	\dot{\bm p}_I = i \int d{\bm x} (\dot{\psi}^* \nabla \psi + \psi^* \nabla \dot{\psi}) = \int d{\bm x} \nabla V({\bm x}) n({\bm x}) ~.
\end{equation}
The first step is a straightforward derivative, and in the second step the equations of motion~(\ref{eq:gpe}) for the condensate wave function $\psi$ were used, and the boundary terms and full derivatives such as $\int d{\bm x} \nabla [n({\bm x})^2]$ do not contribute and can be omitted. Here $n({\bm x}) = |\psi({\bm x})|^2$ is the total density of the condensate. If one considers a fluctuation $\delta \rho({\bm x})$ above symmetric initial state $n_0({\bm x})$ i.e. $n({\bm x}) = n_0({\bm x}) + \delta \rho({\bm x})$, then the symmetric part also does not contribute to the integral because of antisymmetry of $\nabla V({\bm x})$, and the equation finally reads $\dot{\bm p}_I = \int d{\bm x} \nabla V({\bm x}) \delta \rho({\bm x})$, i.e. the Eq.~(\ref{eq.dpdt}) in the main text.

%%%%%%%%%%%%%%%%%%%%%%%%%%%%

\end{document}